\documentstyle[11pt,newpasp,twoside,epsfig]{article}
\newcommand{\magn}[2]{$#1^{\mathrm{m}}\!\!.#2$}

\begin{document}
\title{NGC 6251 at multiple scales and wavelengths}
\author{P. N. Werner, D. M. Worrall \& M. Birkinshaw}
\affil{University of Bristol, Tyndall Avenue, Bristol BS8 4EJ, U. K.}

\begin{abstract}

We have studied the FR I radio galaxy NGC 6251 and its environment at
several wavelengths and scale lengths. On the large scale, we have
probed the gravity field by measuring the velocity dispersion of the
cluster members associated with NGC~6251 and relating this to the
cluster's X-ray emission.  On the small scale, the gravitational
information is provided by cold HI near the nucleus and the
distribution of stars and gas near the centre of the galaxy.  The cold
HI gas which we have measured explains the absorption of the central
X-ray emission and is consistent with the extinction through the
recently discovered HST gas disc of NGC~6251.

\end{abstract}

\section{Introduction}

NGC~6251 is an $m_{B}$ = \magn{13}{6} giant E2 radio galaxy known for
its remarkable radio jet, which is aligned (within a few degrees) from
pc to Mpc scales.  Arguments that the galaxy contains a supermassive
black hole with mass $\sim 10^{9} M_{\odot}$ (Young et al.\,1979) were
recently supported by HST observations of the nuclear gas and dust
disc on a scale of a few $\times100$ pc (Ferrarese \&\ Ford 1999).
NGC~6251 lies within the confines of the Zwicky cluster
Zw~1609.0+8212.

\section{The optical view}

It is believed that the X-ray-emitting gas detected in clusters is in
equilibrium within the gravitational potential of the cluster.  This
means that there is a strong correlation between the X-ray gas
temperature and the velocity dispersion of the  member
galaxies.  This relation has been empirically parametrised by Girardi
et al.\,(1996) as:

\begin{equation}
(\sigma_{z}/\rm{km~s}^{-1})=10^{(2.53\pm0.04)}\times(\it{T}/\rm{keV})^{(0.61\pm0.05)}
\end{equation}

We observed NGC 6251 and 12 nearby Zw1609.0+8212 galaxies (2 of which
have double nuclei) with the Multiple Mirror Telescope (MMT) at
10~\AA\ resolution (Werner, Worrall, \& Birkinshaw 2000).  We measured
redshifts for all the sources in our sample using the standard
cross­-correlation technique of the stellar absorption­-line spectrum
with a template spectrum (Tonry \& Davis 1979).  In half of the cases,
we were able to check the redshifts by fitting gaussians to strong
emission lines.

Redshifts and projected distances from NGC~6251 enable us to rule out
cluster membership for 4 galaxies, leaving 9 members of a cluster with
systemic velocity $V=7328\pm105$ km~s$^{­-1}$ and line-­of-­sight
velocity dispersion $\sigma_{z} = 283\,(+109,-­52)$ km~s$^{-­1}$ .
Inserting this value into equation 1 leads to an estimated X­-ray
temperature of $T=0.7\,(+0.6,-­0.2)$ keV for the cluster's gaseous
atmosphere.

\section{The X-ray view}

\it ROSAT \rm PSPC observations found 90\% of the 0.1--­2 keV flux
from NGC~6251 to be unresolved and coincident with the radio core
(Birkinshaw \& Worrall 1993).  The remaining $\sim10\%$ of the flux
appeared to originate from an extended emission component, probably a
gaseous atmosphere with an FWHM of $\sim3$ arcmin (130 kpc; $H_{0} =
50$ km~s$^{­-1}$ Mpc$^{­-1}$). The limited number of counts in
extended emission did not allow a useful gas temperature measurement
to be made, but we have now been able to estimate it using the
$\sigma_{z}/T$ relation (see section 2).

Birkinshaw \& Worrall (1993) measured the external pressure exerted on
the radio jet by the intra--cluster medium and compared it to the
minimum internal jet pressures calculated by Perley, Bridle, \& Willis
(1984).  They found that the various sections of the jet require gas
temperatures of $T\ga2-5$ keV to confine them.  Such high temperatures
are ruled out by the optical data using the correlation of equation 1,
so the problem of the over-pressured jet persists in NGC~6251.

Combined ASCA and ROSAT spectroscopy of the core favour a steep X-­ray
spectral index and a line­-of-­sight column density within NGC~6251 of
$\sim10^{21}$\ cm$^{-­2}$.  This is consistent with the column density
of $(1.3\pm0.3) \times 10^{21}$ cm$^{-­2}$ through the HST disc
derived from Ferrarese \& Ford's (1999) visual extinction of $A_{V} =
0.61 \pm 0.12$ mag.

\section{The radio view}

We made L­-band spectral-­line measurements of NGC~6251 with the VLA
A­-array to probe for cold gas which would be seen in absorption
against the radio core.  For a total of 3.2 hours of on-­source time,
each of the 31 channels had a width of 400 kHz (83 km~s$^{-­1}$),
providing a total bandwidth of 12.5 MHz (2600 km~s$^{­-1}$).

The inner 75\% (in frequency-­space) of the data were averaged to form
a `pseudo­-continuum' or `channel~0' dataset, which was mapped to
produce a high­-dynamic­-range broadband image of NGC~6251 and its
well­-known radio jet (Figure 1).

\begin{figure}[t]
\begin{center}
\psfig{file=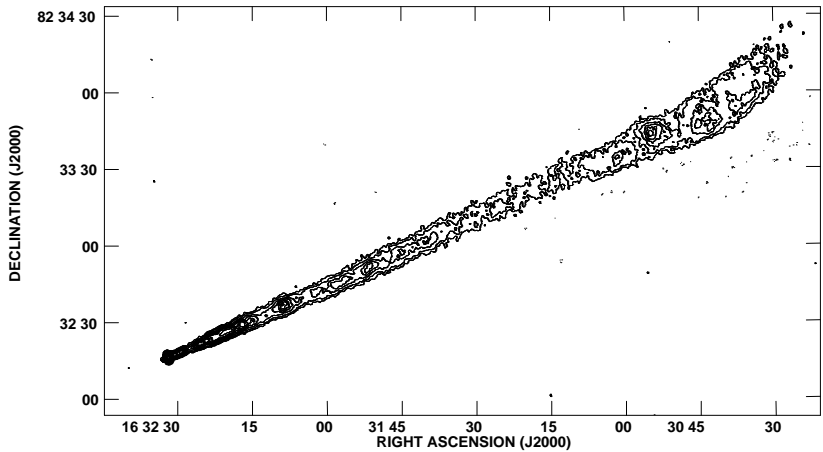,angle=270,width=11cm}

\caption{ ``Pseudo-continuum'' map of the jet of NGC 6251 observed
with the VLA at 21cm. Contours at 0.29 mJy $\times$ (-1, 1, 2, 4, 6,
8, 10, 12, 16, 32, 64, 128, 256, 512, 1024, 2048).  Peak flux is 451
mJy/beam. Scale: 30 arcsec is equivalent to 22 kpc ($H_{0}$ = 50
km~s$^{-1}$ Mpc$^{-1}$).  }

\end{center}
\end{figure}

\begin{figure}[t]
\begin{center}
\epsfig{file=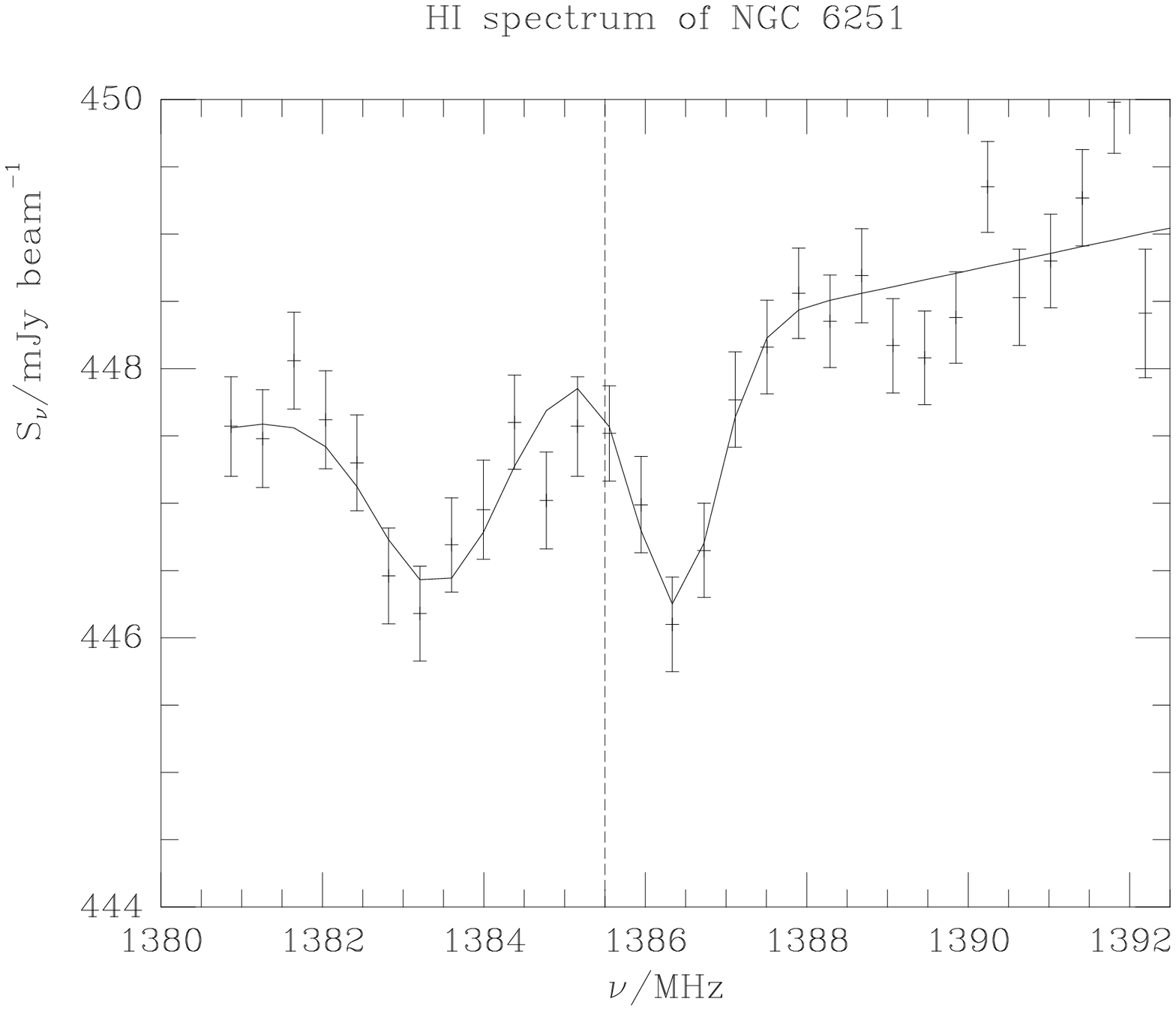,angle=0,width=7cm}

\caption{Spectrum of the core of NGC~6251.  The synchrotron radio
continuum source is absorbed by neutral hydrogen along the line of
sight.  There are two clear absorption features, detected at 
$>5 \sigma$, one on either side of the (optical) redshift (marked by the
dotted line). The fit consists of a baseline, slope, and two
gaussians, and yields $\chi^{2}$ = 29.3 for 22 degrees of freedom.}

\end{center}
\end{figure}

The line data were reduced and a spectrum of the (unresolved) core was
extracted (see Figure 2). Two absorption features are seen against the
synchrotron radio continuum source.  One is sightly redshifted with
respect to the galaxy, and could represent infalling gas fuelling the
active nucleus.  The blue­shifted component may correspond to an
outflow or a cloud of neutral hydrogen orbiting the galaxy with a
strong radial component.  The two features have optical depths of
$\tau_{1386} = 0.0045$ and $\tau_{1383} = 0.0034$, respectively. The
combined column density of the absorbing neutral hydrogen is $4.5
\times 10^{20}$ cm$^{-­2}$ (assuming a spin temperature $T_{s} = 100$
K).

In section 3, it was pointed out that the column densities derived
from X-ray and HST observations were consistent.  To this united
picture we can now add a third quantity -- a direct observation of
cold HI gas absorbed against the radio core, indicating that perhaps
the same material is responsible for the three effects.

\section{Conclusions}

\begin{itemize}

\item There is a poor cluster of galaxies associated with NGC~6251; it
is one of at least three distinct subclusters that may be subunits
of Zw~1609.0+8212.
 
\item The temperature of the cluster atmosphere inferred from the
velocity dispersion measurement is not high enough to provide static
pressure confinement of any part of the kpc­-scale radio jet (at
distances of 5 to 200 kpc from the core).

\item The HI spectrum shows two absorption features, one redshifted
and one blueshifted with respect to the galaxy, with a combined column
density $N_{HI} = 4.5 \times 10^{20} \times (T_{s} /100$ K)
cm$^{-­2}$.

\item The column densities to the nucleus derived from VLA, HST, and
combined ROSAT and ASCA observations are all close, and there appears
to be no need for extra absorption to the central X­-ray or radio
source, over and above that through the HST disc.

\end{itemize}

\end{document}